# A survey of liquid crystalline oxadiazoles

A.Sparavigna
Dipartimento di Fisica, Politecnico di Torino
C.so Duca degli Abruzzi 24, 10122 Torino, Italy
E-mail, amelia.sparavigna@polito.it

**Abstract**
We propose a survey of those oxadiazole compounds, which are mesogenic. We will see that they can have bend-shaped molecules, a biaxial nematic phase, or other interesting peculiarities in the nematic and smectic phase. With large electric dipoles and luminescent properties, these materials are also very appealing for technological applications.

## 1. Introduction

Certain oxadiazoles compounds can reveal liquid crystalline features and in particular, as recently demonstrated, can show the elusive biaxial nematic phase. Theoretically proposed in 1970 by Freiser, the biaxiality can be observed in the nematic phase of boomerang-shaped oxadiazoles [1-3]. In general, the oxadiazoles are interesting for applications to electroluminescent devices where they are the emissive materials [4].

Before starting the survey of recent researches and results obtained with these compounds, let us discuss the basic structure of the oxadiazole molecule. Because it is mainly the oxadiazole group that gives the shape to the final molecule in which it is inserted. Usually, it is explained that liquid crystals are mesomorphic due to their molecular rod-like or disk-like shape. If the structure is bent, the material can produce a spontaneous symmetry breakdown, as the banana-shaped molecules of Niori et al.[5]. The experimental discovery by Tokyo Technology group in 1996, of a ferroelectric behaviour in a liquid crystal phase of achiral bend-core (banana) molecules, represented a revolution for the liquid crystals community. It is then not surprising the interest for any compound, where the shape is not rod-like or disk-like.

The oxadiazole is an aromatic heterocycle molecule, which structure is shown in Fig.1. It has carbon, hydrogen, nitrogen and oxygen. In fact, two nitrogen atoms (for instance, oxazole has one nitrogen atom). There are four possibilities to arrange nitrogen and oxygen, as the figure shows. Moreover, we have the possibility to substitute or di-substitute the hydrogen in the other two positions. An example is the 1,2,5-oxadiazole-3,4-diamine compound in the figure. We well imagine how longer threads attached to the oxadiazole, instead of diamine groups, can create a possible mesogenic compound, with a bend-shape. For this reason, the oxadiazole was used to build the boomerang-shaped molecules.

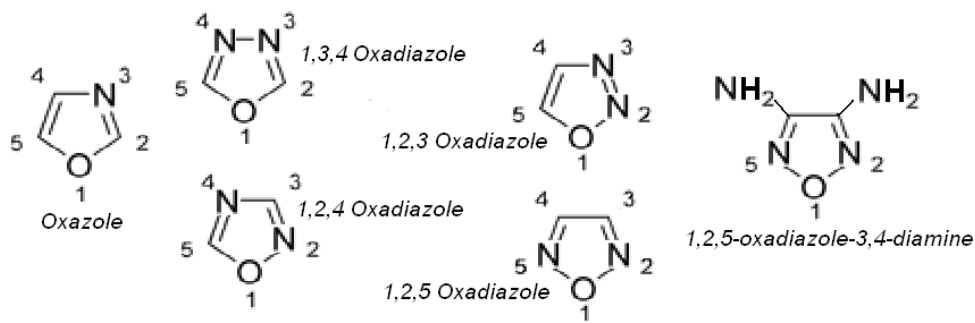

*Fig.1 The four oxadiazoles, the oxazole and an example of a substance with the 1,2,5-oxadiazole.*

**2. Symmetric compounds with 1,3,4-oxadiazoles and biaxiality.**
References [1,2] show the boomerang molecules obtained with the 1,3,4-oxadiazole. These molecules have a biaxial molecular arrangement in the smectic A phase. But they have a biaxial nematic too. This is a mesophase where three distinct optical axes are present, instead of the uniaxial nematic. The uniaxial phase has a single preferred axis, around which the system is rotationally symmetric $D_{\infty h}$. The symmetry group of a biaxial nematic is $D_{2h}$. This group can be represented by the geometry of a rectangular right parallelepiped. And in fact, a route to biaxiality in nematics is the development of molecules, which have the shape of parallelepipeds.

Numerical simulations of transitions from the ordinary nematic to the biaxial one give the evolution of schlieren textures [6], as they can be observed with the polarized light microscope. The problem is more complex than what the common feeling suggests, simply telling that it is the charge of disclinations, which distinguishes the biaxial form from the uniaxial phase. To see beautiful images obtained with polarised microscope of such transition, look at Ref. [7] where mesophases of 1,4"-*p*-terphenyl-bis-2,3,4-tri-dodecyloxy-benzal-imine are investigated. This is a calamitic compound with a biaxial nematic phase. An example of how the texture transition appears, when the nematic passes from uniaxial to biaxial symmetry, inspired from images in [7], is show in Fig.2.

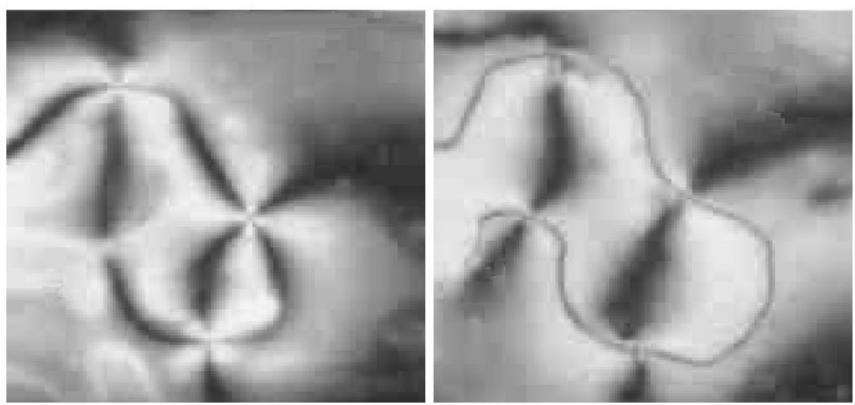

*Fig.2 The transition from the uniaxial (on the left) to the biaxial nematic (on the right) as shown by the schlieren texture.*

In spite of the fact that a biaxial nematic prefers to have a schlieren texture with disclinations of strength 1/2, unambiguous results are given by $H^2$-NMR investigations [1]. The resonance demonstrates that ODBP-Ph-$C_7$ and ODBP-Ph-$C_{12}$ have a truly biaxial nematic (see Fig.3 for the molecular structure). The researchers defined these molecules as boomerangs, because the angle at the core of the molecule, which is the angle determined by the oxadiazole group, is approximately 140°, greater than the typical value of 120° of banana mesogens. There are of course biaxial-shaped calamitic materials, but ODBP boomerangs have a large electric dipole (≈ 4D), which adds the possibility of strong intermolecular association.

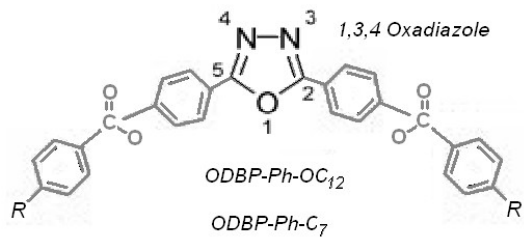

*Fig.3 Boomerang molecules. R is the different fragment $OC_{12}$ or $C_7$.*

**3. The 2,5-disubstituted-1,3,4-oxadiazoles.**
Many compounds with 1,3,4 oxadiazoles have been prepared to study their luminescence property [4]. The compounds are generally chemically stable, sometimes with good electron-transport properties. The study on these materials allows the conclusions that the presence of mesophases is affected by both electronic and steric factors within terminal substituents [4]. Smectic and nematic, enantiotropic and monotropic phases were observed with polarized light microscope. Let us remember that an enantiotropic phase can be observed both heating and cooling the material, whereas a monotropic phase can be entered only during the heating or the cooling, but not both.
In Ref.4 is reported how a large molecular dipole can be obtained inserting a F-substituent in oxadiazoles with 2,5-diphenyl-1,2,3-oxadiazole (electron-deficient) and *p*-alkoxyphenyl (electron-rich) moieties. The F atom, an electron-drawing element, perturbs the charge distribution producing the dipole. The insertion of F atom is stabilising, whereas the insertion of electron-donor substituents destabilise the mesophases. Liquid crystalline oxadiazole compounds with halogens (both 1,3,4-oxadiazoles and 1,2,4-oxadiazoles, see Fig.4) were investigated also in Ref.[8-10].
The importance of oxadiazole compounds is due to the simultaneous presence of a bend-shape and of an electric dipole; this fact is very attractive in the framework of researches on materials with achiral molecules, which could display ferroelectric states. The fact that the dipolar nature of the oxadiazole core is influencing the mesomorphism was shown by molecular simulations too [11].

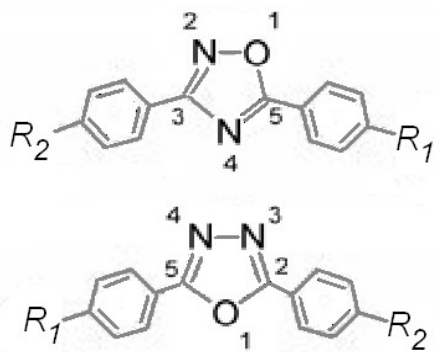

*Fig.4 The ODBPs structures with 1,3,4 and 1,2,4 oxadiazoles.*
*$R_1$ and $R_2$ are substituents with can be halogen atoms.*

## 4. The 3,5-disubstituted-1,2,4 oxadiazoles with bend-shape.

The Liquid Crystal Group of the Organic Intermediates and Dyes Institute in Moscow was the first group to study 3,5-disubstituted 1,2,4-oxadiazoles as fragments of liquid crystalline compounds [8-10]. Before their studies, these compounds were used only for pharmaceutical purposes. In Ref.12, several asymmetric heterocyclic structures are proposed, leading to molecules with bend-shape and mesogenic properties (Fig.5). The 1,2,4-oxadiazole compounds display distinct aspects such as an asymmetrical distribution of electronic density, giving a dipole moment of ≈ 1.5 D. A distortion of the linearity of the molecule due to the bonding angles is observed and then the authors named the compounds as banana-shaped 1,2,4-oxadiazoles.

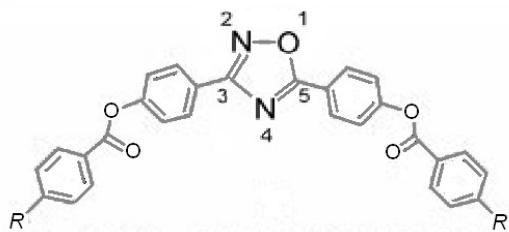

*Fig.5 A banana-shaped compound from Ref.12.*

## 5. Asymmetric oxadiazoles.

The correlation between the existence of mesophases and the chemical structure of molecules is always important for the research in liquid crystals. When a family of asymmetric structures is under investigation, the variety of results is strongly enhanced. This is true for mesomorphic oxadiazole compounds too. These asymmetric materials have been developed [8], because their molecular structure with heterocycles and several possible substituents, gives a variety of smectic and nematic mesophases. The examples we can find in the references that we have previously reported, show that not only the chemical structure of the substituents, but also their position with respect to the oxadiazole ring is relevant for the mesophases.

The optical microscope investigations show very interesting behaviour of the smectic and the nematic phase of some of these oxadiazole compounds [13]. In particular the smectic phases have a toric texture that transform in a spherulitic nematic. Moreover, it is possible to observe a remarkable behaviour in the nematic phase of 3,5-disubstituted-1,2,4-oxadiazoles, a texture transition driven by the temperature. We consider a texture transition as a change in the nematic texture observed by means of the polarised light microscope. The low temperature nematic has a texture with spherulitic domains. Fig.6 shows the transition between the smectic and the spherulitic nematic of compound C of Ref.13 (here simply called ODBP). The figure shows the smectic (a) and the low-temperature (b) and the high-temperature (c) nematic phases.

Texture transitions have been previously observed inside the nematic phase of some mesomorphic thermotropic materials belonging to the families of the alkyloxybenzoic and cyclohexane acids [14–16]. In some cases, peaks in DSC scanning and dielectric spectroscopy investigations accompany the evidence of the texture transition observed by polarised light microscopy. [17]. The Fig.6 in its upper part shows the smectic (a) and the two nematic phases of the alkyloxybenzoic acid HOBA.

Alkyloxybenzoic and cyclohexane acids have mesophases because the mesogenic units are hydrogen-bonded dimers. The more common explanation for the presence of a texture transition in the nematic range is based on the existence of cybotactic clusters of such dimers, favouring a local smectic order in the nematic range [18]. Of course the texture transition inside the nematic phase of oxadiazole compounds can be due to surface effects, but we like to suggest a possible existence of cybotactic clusters in these materials too. In Ref.2, the possibility of cybotactic cluster is also reported. We avoid any suggestion of biaxiality, because it must be checked with further investigations.

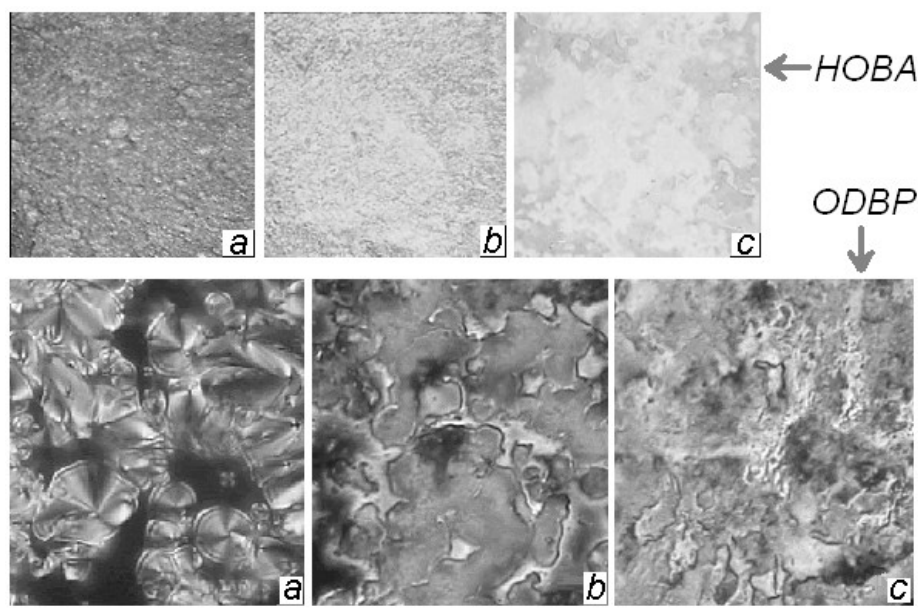

*Fig.5 The smectic and nematic phases of HOBA [16] and ODBP (compound C in [13]). The smectic phase (a) turns in a nematic with smectic-like texture (b). This nematic transforms into an ordinary nematic (c) when subjected to further increase of temperature.*

Hydrogen-bonded oxadiazoles have been also prepared and studied [19], to understand the role of an increased flexibility of molecules. One of the parameters of materials with bend-core molecules, relevant for mesomorphism, is the bend angle. If this angle is too small, the mesomorphism is usually suppressed. We have seen boomerang oxadiazoles, where molecules have 140° for the bend angle, and that display the biaxial phase. If the angle is greater, the behaviour of the system is expected to be that of a calamitic material. The suggestion in [19] is to use flexible linked dimers, to tune the value of the bend angle and see if and when a biaxial or a calamitic behaviour is observed.
Asymmetric molecules containing metals and 1,3,4-oxadiazole and 1,3,4-thiadiazole heterocycles have been proposed [20]. These materials are named as metallomesogens if exhibit thermotropic mesophases and are classified according to the bond formed between the metal and the organic molecule. The interest in synthesising of metal-containing liquid crystals is due to advantages in combining properties of liquid crystal systems with those of transition metals, for applications in optical devices.

## 6. Oxadiazoles for electroluminescence
Organic compounds mostly behave like insulators. But organic compounds with proper structure and orientation of molecules can transport electric current upon certain conditions, competing then with conventional materials used in electronics. Chains of polythiophenylenes for instance, have a charge transport comparable to poor metallic conductors.
In electronics, organic compounds can operate as rectifiers if they have a donor electron-rich part and an acceptor electron-poor part, linked by an insulating bridge. The lowest unoccupied molecular orbital (LUMO) of the acceptor and the highest occupied molecular orbital (HOMO) of the donor are used for transport of electrons. Some organic compounds with proper HOMO and LUMO orbitals can emit light when an electric current is passing through them. This is the electroluminescence. It is based on an electron transfer from LUMO- and HOMO orbitals. The molecule in an excited state comes back to its ground state by electron transfer from LUMO to HOMO orbital, accompanied by irradiation of energy difference in the form of light.

Materials that can simultaneously act as liquid crystals, charge transport agents and emitters are of interest for potential applications in organic light emitting diodes. In 1990, it was found that a compound with 1,3,4-oxadiazole is an excellent electron transport material in an organic electroluminescent diode [21]. After this report, the research to use various oxadiazole molecules to obtain high electroluminescence performances strongly increased. Furthermore, in recent studies of polymer light-emitting diodes, the oxadiazole moieties were demonstrated to possess a high potentiality for transport [22]. The charge drift mobility of oxadiazole derivatives doped in polycarbonate were studied, with the time-of-flight technique, in [23] and discotic oxadiazoles, with columnar mesophases, were proposed for applications in organic electronics [24].

The introduction of oxadiazole moieties to polymer main chains and to mesogens in liquid crystalline compounds is expected to tune the electroluminescence efficiency and the transport properties. Electroluminescent polymers were also prepared, in which thiophene and oxadiazole moieties are connected alternately to form fully conjugated polymers. Likewise, liquid crystalline compounds containing oxadiazole moieties were reported to exhibit a high electron transport capability and a blue electroluminescence emission. [25,26]

**7. Enantioselective segregation and drugs.**

To conclude, let us see another reason to investigate oxadiazoles. Studying the oxadiazoles which display the biaxial nematic phase [27], V. Görtz and J. Goodby found that the achiral biaxial nematic phase can segregate into chiral domains of opposite handedness. In fact then, they observed a nematic phase exhibiting the properties of a conglomerate. Enantioselective segregation has been obtained in the smectic and nematics phases of other banana oxadiazoles, without introducing any chiral species, by a Japanese group [28].

To have such segregations is rather important for drugs. The stereoisomers of chiral drugs often exhibit pronounced differences in their properties, so pronounced that it is necessary to study each stereoisomer separately. This is the reason why the pharmaceutical research has converged to use single enantiomers as substitutes for their racemates. In this framework, new compounds with 1,2,4-oaxdiazoles are developed for medicine production [29]. We mentioned in Sect.4 the fact that 1,2,4-oxadiazoles are significant in terms of their pharmacological properties. In fact, new developments in medicine can start from liquid crystals laboratories with further investigations on boomerang and banana-oxadiazoles.

**References**
[1] L.A. Masden,T.J. Dingemans, M.Nakata and E.T. Samulski, Phys. Rev. Lett. 92 145505 (2004)
[2] T.J. Dingemans and E.T. Samulski, Liquid Crystals 27 131 (2000)
[3] Freiser M.J. Phys. Rev. Lett. 24 1041 (1970)
[4] JieHan, S. Sin-Yin Chui, Chi-Ming Che, Chem. Asian J. 1 814 (2006)
[5] T. Niori, F. Sekine, J. Watanabe, T. Furukawa, and H. Takezoe, J. Mater. Chem., 6, 1231 (1996)
[6] C. Chiccoli, I. Feruli, O. D. Lavrentovich, P. Pasini, S.V. Shiyanovskii, C. Zannoni, Phys. Rev. E 66 030701 (2002).
[7] S. Chandrasekhar, Geetha G. Nair, D. S. Shankar Rao, S. Krishna Prasad, K. Praefcke and D. Blunk, http://www.iisc.ernet.in/currsci/nov251998/articles21.htm

[8] L.A. Karamysheva, S.I. Torgova, I.F. Agafonova and N.M. Shtikov, Mol. Cryst. Liq. Cryst. 160 217 (1995)
[9] O. Francescangeli, L.A. Karamysheva, S.I. Torgova, A. Sparavigna and A. Strigazzi, Proceedings of SPIE 3319 139 (1998)
[10] L.A. Karamysheva, I.F. Agafonova and S.I. Torgova, Mol. Cryst. Liq. Cryst. 332, 407 (1999)
[11] J. Peláez and M.R. Wilson, Phys. Rev. Lett. 97 267801 2006
[12] S.I. Torgova, T.A. Geivandova, O. Francescangeli and A Strigazzi, Pramana. 61(2) 239 (2003)
[13] A. Sparavigna, A. Mello and B. Montrucchio, Phase Trans. 80 987 (2007)
[14] A. Sparavigna, A. Mello and B. Montrucchio, Phase Trans. 79 293 (2006); Phase Trans, 80 191 (2007).
[15] M. Petrov, A. Braslau, A.M. Levelut and G. Durand, J. Phys. II (France) 2 1159 (1992)
[16] P. Simova and M. Petrov, Phys. Stat. Sol. A 80 K153 (1983); M. Petrov and P. Simova, J. Phys. D: Appl. Phys. 18 239 (1985). [10] B. Montrucchio, A. Sparavigna and A. Strigazzi, Liq. Cryst. 24 841 (1998); B. Montrucchio, A. Sparavigna, S.I. Torgova and A.Strigazzi, Liq. Cryst. 25 613 (1998).
[17] L. Frunza, S. Frunza, A.Sparavigna, M. Petrov and S.I. Torgova., Mol. Mater. 6 215 (1996).
[18] A. De Vries, Mol. Cryst. Liq. Cryst. 10 31 (1970).
[19] P.J. Martin and D.W. Bruce, Liq. Cryst. 34 767 (2007)
[20] M. Parra, S. Hernández and J. Alderete, J. Chil. Chem. Soc. 48 on line (2003)
[21] C. Adachi, T. Tsutsui and S. Saito, Appl. Phys. Lett. 56 799 (1990)
[22] C. Hosokawa, N. Kawasaki, S. Sakamoto and T. Kuzumoto, Appl. Phys. Lett. 62 2503 (1992)
[23] H. Tokuhisa, M. Era, T. Tsutsui, and S. Saito, Appl. Phys. Lett. 66 3433 (1995)
[24] Y-D. Zhang, K.G. Jespersen, M. Kempe, J.A. Kornfield, S. Barlow, B. Kippelen and S.R. Marder, Langmuir 19 6534 (2003)
[25] H Mochizuki, T. Hasui, M. Kawamoto, T. Shiono, T. Ikeda, C. Adachi, Y. Taniguchi and Y. Shirota, Chem. Commun. pp.1923–1924 (2000)
[26] Fan R. , Culjkovic D. , Piromreun P.; Turon M.J.; Langseth J.E. , Milliaras G.G., Gu Shihai, Sukhomlinova L., Twieg R.J. , Proc. SPIE, Organic Light-Emitting Materials and Devices III, Zakya H. Kafafi Ed. 3797 170 (1999)
[27] V. Görtz  and J.W. Goodby,  Chem Commun. pp.3262-3264 (2005)
[28] S-W. Choi, S. Kang, Y. Takanishi, K. Ishikawa, J. Watanabe, H. Takezoe, Chirality 19  250 (2007);  Chirality 19 519 (2007)
[29] V. Martins, L. Braga, S.J. de Melo, R.M. Srivastava, E.P. da S. Falcão, J. Braz. Chem. Soc. 15 on line (2004)